\documentclass[article,10pt,superscriptaddress,onecolumn,floatfix,showpacs,superscriptaddress,nofootinbib]{revtex4-2}
\usepackage{graphicx}
\usepackage{subfigure}
\usepackage{bm}
\usepackage[bbgreekl]{mathbbol}
\usepackage{amsmath} 
\let\oldAA\AA
\renewcommand{\AA}{\text{\normalfont\oldAA}}

\usepackage{lipsum, babel}
\usepackage{color}
\usepackage{verbatim}

\def\Eb{{\bf E}}
\def\g2R{g^{(2)}_R}

\def\w0{w_0}

\def\GG{{\bf G}}

\def\db{{\bf d}}  
\def\Eb{{\bf E}}  
\def\rb{{\bf r}}

\def\hge{\hat{\sigma}^{ge}}  
\def\heg{\hat{\sigma}^{eg}}

\def\db{\textbf{d}}

\def\ket#1{\mathinner{|{#1}\rangle}}
\def\db{\boldsymbol{\wp}}  
\def\dbu{\hat{\boldsymbol{\wp}}}

\def\db{\boldsymbol{\wp}}  
\def\dbu{\hat{\boldsymbol{\wp}}}

\def\Heff{\hat{H}_{\rm eff}}

\newcommand{\fref}[1]{Fig.~\ref{#1}}

\begin{document}
\title{Optical Properties of Concentric Nanorings of Quantum Emitters}
\author{Verena Scheil}
\affiliation{Institut f\"ur Theoretische Physik, Universit\"at Innsbruck, Technikerstr. 21a, A-6020 Innsbruck, Austria}
\author{Raphael Holzinger}
\affiliation{Institut f\"ur Theoretische Physik, Universit\"at Innsbruck, Technikerstr. 21a, A-6020 Innsbruck, Austria}
\author{Maria Moreno-Cardoner}
\affiliation{Departament de F\'{i}sica Qu\`{a}ntica i Astrof\'{i}sica and Institut de Ci\`{e}ncies del Cosmos,\
Universitat\ de\ Barcelona,\ Mart\'{i}\ i\ Franqu\`{e}s\ 1, E-08028\ Barcelona,\ Spain.}
\affiliation{Institut f\"ur Theoretische Physik, Universit\"at Innsbruck, Technikerstr. 21a, A-6020 Innsbruck, Austria}
\author{Helmut Ritsch}
\affiliation{Institut f\"ur Theoretische Physik, Universit\"at Innsbruck, Technikerstr. 21a, A-6020 Innsbruck, Austria}
\date{\today}
%\email{xx@xx}
%\address{}

\date{\today}
\begin{abstract}
A ring of sub-wavelength spaced dipole-coupled quantum emitters features extraordinary optical properties when compared to a one-dimensional chain or a random collection of emitters. One finds the emergence of extremely subradiant collective eigenmodes similar to an optical resonator, which feature strong 3D sub-wavelength field confinement. Motivated by structures commonly appearing in natural light harvesting complexes, we extend these studies to stacked concentric multi-ring geometries. We predict that double rings allow to engineer significantly darker and better confined collective excitation states over a broader energy band compared to the single ring case. These potentially enhance weak field absorption and low loss excitation energy transport. For the specific geometry of the three rings appearing in the natural LH2 antenna we show that the coupling between the lower double ring structure and the higher energy blue shifted single ring is very close to a critical value for the actual size of the molecule. This creates collective excitations with significant contributions from all three rings as a vital ingredient for efficient and fast coherent inter-ring transport. Such a geometry thus should prove useful for the design of sub-wavelength weak field antennas.
\end{abstract}
\maketitle
\section{Introduction}
The optical properties of a quantum emitter, such as its excitation lifetime and transition frequency, are strongly modified when it is placed close to a second emitter, due to vacuum fluctuations that mediate dipole-dipole interactions between them. As a remarkable example, the decay rate of a collection of emitters separated by subwavelength distances can be enhanced or suppressed, leading to the well known phenomena of superradiance or subradiance, respectively  \cite{dicke1954coherence,haroche1982superradiance, guerin2016subradiance,Solano2017superradiance}. These phenomena are expected to be strongly enhanced in ordered subwavelength arrays of emitters, where maximal interference of the scattered fields can be observed \cite{Porras2008collective,zoubi2008bright,Jenkins2012Controlled,Jenkins2013metamaterial,plankensteiner2015selective,bettles2015cooperative,Tudela2015Subwave,sutherland2016collective,bettles2016cooperative,bettles2016enhanced,shahmoon2016cooperativity,asenjo2016,asenjo2017exponential,ruostekoski2017arrays,hebenstreit2017subradiance,Chang2018colloquium,guimond2019subradiant,PineiroOrioli2019dark,zhang2019theory, kornovan2019extremely,Zhang2020subradiant,Zhang2020universal,PineiroOrioli2020subradiance}. 

Among the different array geometries, a ring-shaped structure formed by regularly placed emitters has very special optical properties. It has been shown before \cite{zoubi2008bright,plankensteiner2015selective,asenjo2017exponential}, that a linear chain of emitters whose inter-particle distance is smaller than half of the light wavelength supports collective modes that can guide light and are extremely subradiant with the excitation lifetime increasing polynomially with the atom number. The lifetime limitation arises from photon scattering off the ends of the chain. Remarkably, by joining the ends of the chain to form a closed ring, the lifetime can be exponentially increased with atom number \cite{asenjo2017exponential,moreno2019subradiance,cremer2020polarization}. 

Such extraordinary optical properties can be exploited for applications including efficient energy transfer, single photon sources or light harvesting \cite{Mattioni2021,Mattiotti2022}. We have previously shown  \cite{moreno2019subradiance,cremer2020polarization}, that tailoring the geometry, orientation and distance between two such nanorings allows for lossless and high fidelity transport of subradiant excitations, as if the two rings were two coupled nano-scale ring resonators. Besides subradiant states confining and guiding light, these nanorings also feature radiant modes whose corresponding electromagnetic field is strongly focused at its center. By placing an extra emitter at its center, these modes can be exploited to create a nano-scale coherent light source with a spectral line width which is strongly suppressed compared to the single atom decay rate \cite{holzinger2020nanoscale}. In this case the collective optical modes of the ring play the role of the cavity modes and the central atom acts as the gain medium when incoherently pumped. Furthermore, if the central emitter is absorptive, the system can be tailored to achieve a strong absorption cross section way beyond the single atom case, while the outer ring behaves as a parabolic mirror when illuminated externally by a coherent light field \cite{moreno2022efficient}.

In this work, we analyse in detail how the optical properties of two or more of these nanorings are modified when they are stacked in a concentric way. Note that this system is radically different compared to the case previously studied of two rings coupled side by side \cite{moreno2019subradiance,cremer2020polarization}, as it preserves some rotational symmetry. The study of this geometry is strongly motivated by the abundant presence in nature of highly efficient photosynthetic complexes sharing a similar stacked structure \cite{McDermott1995Crystal,bourne2019structure}. In particular, the active core photosynthetic apparatus of certain bacteria is formed by chromophores, featuring an optical dipole transition, which are arranged symmetrically forming a complex structure of stacked concentric coupled nanorings. Some of these units are specialized in transforming the absorbed energy into chemical energy (LH1), while a larger number of them (LH2 and LH3) do not have a reaction center but efficiently capture and funnel light towards the LH1 units. 

In this system, coherence effects between the chromophores have already shown to play a crucial role in the energy transfer and light harvesting \cite{Jang2004Multi,Jang2013Resonance,Plenio}. A natural question is whether collective decay, i.e., superradiance and subradiance, plays an essential role in this process, and whether nature chooses a particular geometry in order to optimize its effects. In this work, we aim at shedding light on this question, by analyzing the optical properties and exciton dynamics in realistic structures. Furthermore, similar mechanisms could be in principle exploited for artificial light harvesting. Proving these concepts could be already possible using state-of-the-art experimental setups, such as neutral atoms trapped in optical lattices \cite{rui2020subradiant,bakr2009quantumgas,sherson2010single,weitenberg2011single}, optical tweezer arrays \cite{barredo2016Atom,Endres2016,Barredo2018,Norcia2018,Schlosser2019Large}, microwave coupled superconducting q-bits \cite{wang2020controllable,mirhosseini2019Cavity,vanLoo2013Photon} or solid-state quantum dots \cite{fontcuberta2017engineering,fontcuberta2020measuring}.

The paper is organized as follows. We first introduce the theoretical framework to describe a system of dipole-dipole interacting quantum emitters, and demonstrate that a structure of coupled symmetric nanorings can be described in a particularly simple form in terms of Bloch eigenmodes. Next, we summarize the optical properties of single nanorings, which can exhibit special radiating properties. We then move to study the case of two coupled nanorings, displaying two energy bands. Thereafter, we apply a similar analysis to elucidate the radiating properties of a realistic natural light-harvesting complex (LH2), which contains a closely double ring structure with a shifted third ring at higher resonance frequencies. Studying this geometry we find that the rings geometry and size is critically close to the case where the energy bands of all rings overlap to for common superradiant exciton states. 

\section{Theoretical Framework: Bloch Eigenmodes}

\begin{figure*}
\centering
\includegraphics[width=0.8\textwidth]{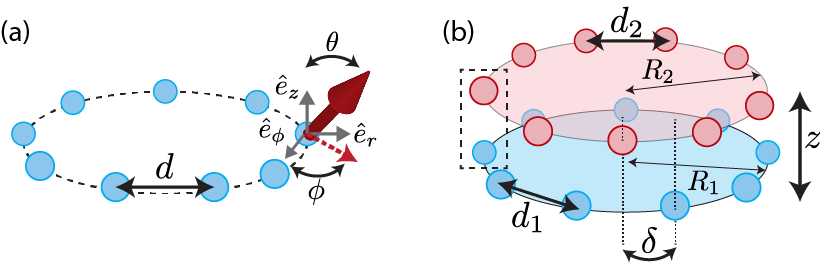}
\caption{{\bf (a)} Schematics of a single ring with lattice constant $d$. Each emitter features an optical dipole moment (indicated by the red solid arrow) with orientation $\dbu = \sin\theta\cos\phi \, \hat{e}_{\phi} + \sin\theta\sin\phi \, \hat{e}_{r}+\cos\theta \,\hat{e}_z$, where $\theta$ and $\phi$ are the polar and azimuth angle, respectively. The vertical, radial and tangential unit vectors are indicated by $\hat{e}_z$, $\hat{e}_r$ and $\hat{e}_\theta$, respectively. The red dashed arrow denotes the projection of the dipole onto the ring plane. {\bf (b)} Double ring structure: two rings of radius $R_1$ and $R_2$ and lattice constants $d_1$ and $d_2$ are stacked concentrically and separated by the vertical distance $z$. The two rings are in general rotated by an angle $\delta$. The dashed-line  rectangle encloses the two sites (one from each of the rings) forming a possible unit cell (see main text).}
\label{fig1}
\end{figure*}

Let us consider first a ring-shaped array (or regular polygon) of $N$ identical two-level quantum emitters with minimum inter-particle distance $d$. The emitters possess a single narrow optical dipole transition around the frequency $\omega_0$ with dipole orientation $\dbu_i = \sin\theta\cos\phi \, \hat{e}_{\phi,i} + \sin\theta\sin\phi \, \hat{e}_{r,i}+\cos\theta \,\hat{e}_z$ ($i=1,\dots, N$), where $\hat{e}_z$ and $\hat{e}_{r,i (\phi,i)}$ denote unit vectors along the vertical and radial (tangential) direction defined with respect to the emitter $i$, respectively  [see \fref{fig1}(a)]. In this work we will then consider a configuration where two or more of these rings are stacked concentrically around the $\hat{z}$-axis [see \fref{fig1}(b)]. 

All the emitters are dipole-dipole interacting via the electromagnetic field vacuum fluctuations. After integrating out the optical degrees of freedom in the Born-Markov approximation~\cite{lehmberg1970radiation}, the atomic reduced density matrix is governed by the master equation $\dot{\rho} = -i[ H ,\rho ] + \mathcal{L}[\rho]$ ($\hbar \equiv 1$), with the dipole-dipole Hamiltonian 
\begin{equation}
  H = \sum_{ij;i \neq j} \Omega_{ij} \hge_i \heg_j,
  \label{eq:Hamiltonian}
\end{equation}
and Lindblad operator
\begin{equation}
  \mathcal{L}[\rho] = \frac{1}{2} \sum_{i,j} \Gamma_{ij} \left( 2\hge_i \rho \heg_j - \heg_i \hge_j \rho - \rho \heg_i \hge_j \right),
  \label{eq:Lindblad}
\end{equation}
with $i$ and $j$ running over all dipoles. The coherent $\Omega_{ij}$ and dissipative $\Gamma_{ij}$ dipole-dipole couplings can be written in terms of the Green's tensor $\GG(\rb,\omega_0)$ in free space:
\begin{align}
\Omega_{ij} &= - \frac{3\pi \Gamma_0}{k_0} ~\textrm{Re}\left\{\dbu_i^*\cdot \GG (\rb_i -\rb_j,\omega_0) \cdot \dbu_j\right\},
\\
\Gamma_{ij} &= \frac{6\pi \Gamma_0}{k_0} ~\textrm{Im}\left\{\dbu_i^* \cdot \GG (\rb_i -\rb_j, \omega_0) \cdot \dbu_j\right\},
\end{align}
where $\rb_i$ is the position of the $i$-th dipole and  $\GG(\rb,\omega_0)$ is given by
\begin{align}
    \GG(\rb,\omega_0) &= \frac{e^{i k_0 r}}{4\pi k_0^2r^3} \left[(k_0^2 r^2 + i k_0 r -1) \mathcal{I} - (k_0^2 r^2 +3i k_0r -3)\frac{\rb \otimes \rb^{\rm T}}{r^2}\right] .
\end{align}

Here, $k_0 = \omega_0 /c = 2\pi/\lambda$ is the wavenumber associated with the atomic transition, $\lambda$ the transition wavelength, and $\Gamma_0 = \left| \db \right|^2 k_0 ^3 / 3 \pi \epsilon_0$ is the decay rate of a single emitter with dipole moment strength $|\db|$.

The scattered electromagnetic field can be also retrieved from a generalized input-output relation~\cite{asenjo2016,asenjo2017exponential} once the atomic coherences are known:
\begin{align}
\Eb^+(\rb) = \frac{|\db| k_0^2}{\epsilon_0} \sum_i \GG(\rb-\rb_i,\omega_0) \cdot \dbu_i  \hge_i.
\label{Eq:Fields}
\end{align}

Motivated by realistic conditions in natural light harvesting complexes, this work focuses on the linear optical properties and response of the system at low light conditions. Therefore, we will restrict our study to the case where at most a single excitation is present in the system. In this situation, the first term in the Lindblad operator Eq.(\ref{eq:Lindblad}) (also known as recycling term) only modifies the ground state population and it is not relevant for the observables of interest (e.g. scattered fields or excitation population). The remaining terms in the equation can be recast as an effective non-Hermitian Hamiltonian:
\begin{align}
H_{\rm eff} = \sum_{ij}  \left( \Omega_{ij} -i\frac{\Gamma_{ij}}{2} \right) \heg_i \hge_j.
\label{Heff}
\end{align}
with $\Omega_{ii} = 0$. In this situation, the dynamics of the system can then be fully understood in terms of the collective modes defined by the eigenstates of $H_{\rm eff}$. Each of these modes have associated a complex eigenvalue, whose real and imaginary parts correspond to the frequency shift and decay rate of the collective mode, respectively. As we will see next, for a symmetric ring-shaped structure these modes have a particularly simple form as they correspond to Bloch functions. 

\subsection*{Bloch Eigenmodes in Rotationally Symmetric Ring Structures}
We will consider here ring structures possessing an $N-$fold rotational symmetry, similarly to those arising in certain natural light harvesting complexes \cite{McDermott1995Crystal,bourne2019structure}. In this case, as we will see, the eigenmodes corresponding to the single excitation manifold will be of the Bloch form, i.e., delocalized states with well defined angular momentum $m$.

The $N$-fold rotational symmetry enables defining $N$ different unit cells (for an example, see \fref{fig1}), which will be denoted by $j=1,\cdots,N$. Each cell contains in general $d$ dipoles with given orientations $\dbu_{j\alpha}$ with $\alpha = 1,\cdots,d$. We can then rewrite Eq.\eqref{Heff} as 
\begin{align}
H_{\rm eff} = \sum_{i,j=1}^N \sum_{\alpha,\beta = 1}^d  G_{ij}^{\alpha\beta} \heg_{i\alpha} \hge_{j\beta},
\label{HeffCoupled}
\end{align}
with $G_{ij}^{\alpha\beta} \equiv \dbu_{i\alpha}^* \cdot \GG(\rb_{i\alpha}-\rb_{j\beta}) \cdot \dbu_{j\beta}$.
We note that a structure consisting of several coupled concentric rings with the same emitter number, each ring being rotationally symmetric, can also be described within this model. In this case, the unit cell contains one site of each of the rings, and it has as many components as rings are.

In the following, we demonstrate that the eigenmodes of the coupled structure are of the Bloch form. The symmetry of the system imposes that the position and polarization vectors associated with dipole $i\alpha$ transform under a rotation $\mathcal{U}$ of angle $2\pi/N$ (around the $\hat{z}-$axis) according to $\rb_{i\alpha} \rightarrow \mathcal{U} \rb_{i\alpha} = \rb_{i+1\alpha}$ and $\dbu_{i\alpha} \rightarrow \mathcal{U}\dbu_{i\alpha} = \dbu_{i+1\alpha}$. By noting that $\GG$ is a tensor containing terms proportional to the identity and to $\rb_{i\alpha} \otimes \rb^{\rm T}_{j\beta}$, and thus it transforms under the same rotation as $\GG(\rb_{i\alpha}-\rb_{j\beta}) \rightarrow \mathcal{U} \GG(\rb_{i\alpha}-\rb_{j\beta}) \mathcal{U}^\dagger = \GG(\rb_{i+1,\alpha}-\rb_{j+1,\beta})$, we can then conclude that $G_{ij}^{\alpha \beta} = G_{i+1,j+1}^{\alpha\beta}$. Thus, this coupling matrix can be relabelled as $G_{i+1,j+1}^{\alpha\beta} \equiv G_{\ell}^{\alpha\beta}$, with $\ell = j-i$ ($\ell = 0,\cdots,N-1$), as it is a periodic function only depending on the difference between the two indices $i$ and $j$. This property allows to write the Hamiltonian Eq.\eqref{HeffCoupled} in terms of Bloch modes as follows:
\begin{align}
H_{\rm eff} &= \sum_{i}^N \sum_{\ell = 0}^{N-1} \sum_{\alpha,\beta = 1}^d  G_{\ell}^{\alpha\beta} \heg_{i\alpha} \hge_{i+\ell,\beta} \notag \\
&= \sum_m \sum_{\alpha,\beta = 1}^d  \tilde{G}_m^{\alpha\beta} \heg_{m\alpha} \hge_{m\beta},
\label{HeffDiagonal}
\end{align}
where $\tilde{G}_m^{\alpha\beta} \equiv \sum_{\ell = 0}^{N-1} e^{i2\pi m\ell/N} G_{\ell}^{\alpha\beta}$, and we have defined the creation and annihilation operators of a collective Bloch mode with well defined angular momentum $m$: 
\[\hat{\sigma}^{eg(ge)}_{m\alpha} = \frac{1}{\sqrt{N}} \sum_{\ell=0}^{N-1} e^{(-) i 2\pi m\ell/N} \hat{\sigma}^{eg(ge)}_{\ell\alpha}.\] 
Here, the periodicity of the wavefunction under a $2\pi$ rotation imposes $m$ to be an integer value, and thus, $N$ linearly independent eigenstates can be constructed by choosing $m = 0, \pm 1,\pm 2, \cdots, \lceil \pm (N-1)/2 \rceil$, where $\lceil \cdot \rceil$ is the ceiling function. 

Eq.\eqref{HeffDiagonal} is not yet in its full diagonal form (except if the unit cell contains a single dipole), but it already tells us that the angular momentum is a good quantum number. For each value of $m$, the eigenmodes consist in general of a superposition of each excited dipole in the unit cell and it can be easily found by diagonalizing the $d \times d$ complex $\tilde{G}^{\alpha\beta}_m$ matrix, leading to $\Heff = \sum_{m,\lambda} \left(\Omega_{m\lambda}-i\Gamma_{m\lambda}/2\right)  \hat{\sigma}^{eg}_{m\lambda} \hat{\sigma}^{ge}_{m\lambda}$. Here, $\Omega_{m\lambda}$ ($\Gamma_{m\lambda}$) is the real (imaginary) part of the eigenvalue associated with Bloch mode $m$ and $\lambda$, whereas $\hat{\sigma}^{eg}_{m\lambda}$ is the corresponding creation operator.

\section{Optical Properties of Nanorings}
\subsection{Single Nanoring}
Let us first summarize some of the most relevant optical properties for a single ring with $N$ dipoles, i.e., the case where the unit cell contains just a single dipole. As previously shown in \cite{moreno2019subradiance,cremer2020polarization} the optical properties of the ring strongly depend on the size of the ring compared to the light wavelength and on the dipole orientations. In the following, we focus on two different limiting regimes: a dense large ring (quasi linear chain) and a small ring (Dicke limit).\\

{\bf Dense and large ring case (quasi-linear chain limit).--} A large ring with a large number of emitters locally resembles a linear array, and can support optical modes which do not propagate into the three-dimensional space but are rather confined and guided through the array. These modes correspond to spin-waves (Bloch modes) whose quasi-momentum along the chain is larger than the light wavenumber $k_0$. This leads to an evanescent field along the transverse directions to the array. In the very large ring case, one can identify the linear momentum $k_z \leftrightarrow 2\pi m/N d$, and the condition $k_z>k_0$ sets the value of the angular momentum of the guided subradiant modes to be $m \gtrsim m_0$, with $m_0 = Nd/\lambda$ associated with the light line. Moreover, such states can only exist if $d < \lambda/2$, as the maximum value of $k_z$ (or equivalently $m$) is given by the boundary of the first Brillouin zone. 

Despite these similarities, a striking difference between linear and closed ring configurations is the scaling of the subradiant decay rates with emitter number. Indeed, by closing the ends of the open chain in a ring structure losses can be strongly reduced, leading to an exponential suppression of the decay rates with atom number, in contrast to the polynomial suppression for the linear chain.

\begin{figure*}[ht!]
\centering
\includegraphics[width=1\textwidth]{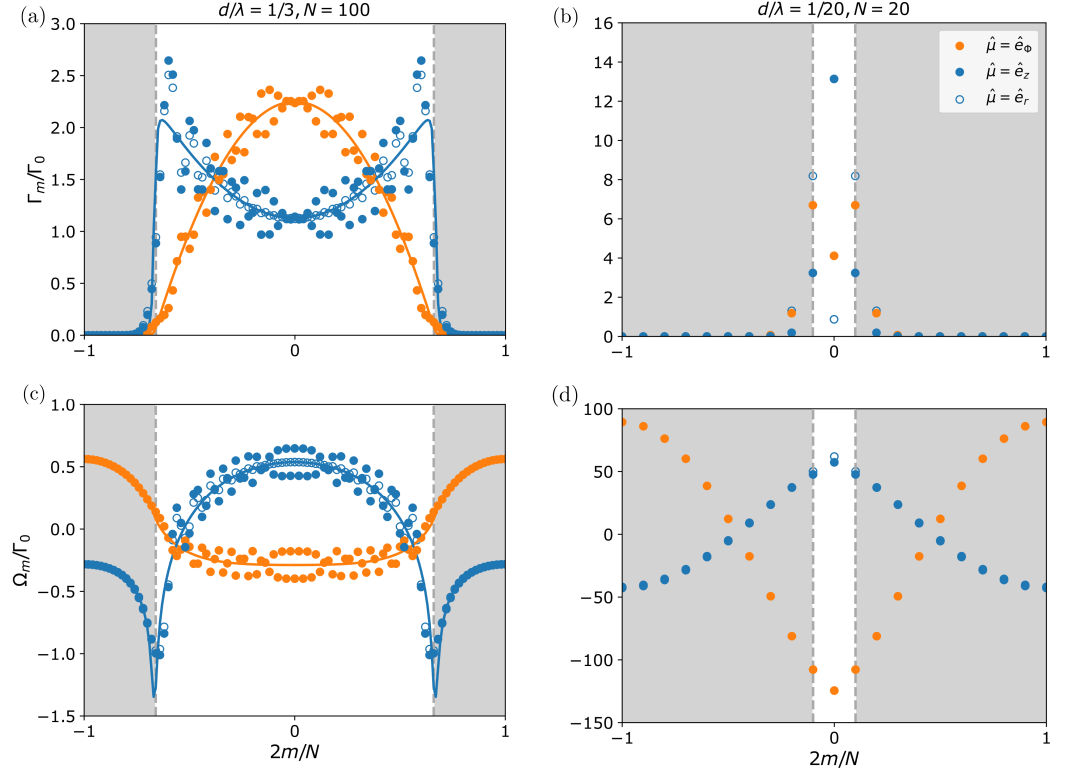}
\caption{{\bf Single Ring Optical Properties}. {\bf(a)}-{\bf(b)} Collective decay rates $\Gamma_m$ and {\bf(c)}-{\bf(d)} frequency shifts $\Omega_m$ versus angular momentum $m$, depending on polarization orientation (blue open, blue solid and orange are for transverse, radial and tangential polarization, respectively). Left panels correspond to a large ring with $d/\lambda = 1/3$ and $N=100$. For comparison, solid lines show the result for an infinite linear chain with transverse (blue) and longitudinal (orange) polarization. Right panels are for $d/\lambda = 0.05$ and $N=20$ (Dicke regime). In this case there are only one (two) bright modes at $m=0$ ($m=\pm 1$) for transverse (tangential and radial) polarization. For tangential polarization the bright (dark) modes are energetically low (high), whereas the opposite behavior is found for radial and transverse polarization. }
\label{Fig2}
\end{figure*}

On the other hand, the modes for which $m \lesssim m_0$ are in general radiant. The angular momentum of the brightest state however, strongly depends on the polarization direction of the atoms. In \fref{Fig2}(a) we have plotted the collective decay rates versus $m$, for a ring of $N=100$ emitters and different polarization orientations $\dbu_i = \left\{\hat{e}_z, \hat{e}_{r,i}, \hat{e}_{\phi,i} \right\}$. For comparison, we also plot the result for an infinitely long linear chain with the same lattice constant (solid line). Clearly, in this regime, the radial and transverse (tangential) polarization decay rates tend to those for the perpendicularly (longitudinally) polarized linear chain, with maximally bright modes close to the light line $m = m_0$ ($m =0$). 

Besides studying the radiative properties, it is also interesting to analyse the sign of the frequency shifts in the collective modes arising due to dipole-dipole interactions. Figure \ref{Fig2}(c) shows the frequency shifts corresponding to  \fref{Fig2}(a). We find that the symmetric $m=0$ mode has a positive (negative) shift when the dipoles are aligned transversely (longitudinally). This is not so surprising when thinking of interacting classical static dipoles which repel (attract) each other if they are aligned in parallel (in a head-to-tail configuration). Note also that in this regime the bright states are always energetically lower than the guided subradiant modes.\\
{\bf Small ring case (Dicke limit).--} We now focus on a different regime where the ring diameter is small compared to the light wavelength, i.e., $R \lesssim \lambda/2$ (Dicke limit). This regime will be relevant in the study of natural light harvesting complexes, given the small  inter-particle distances which are few orders of magnitude smaller than the light wavelength. In this case, the emitters radiate as if they were a single dipole with effective moment strength and decay rate
\begin{align}
\hat{\boldsymbol{\wp}}_{m,\rm eff} = N^{-1/2} \sum_\ell e^{i 2\pi m \ell/N}\hat{\boldsymbol{\wp}}_\ell, \qquad{} \qquad{}
\Gamma_m = |\hat{\boldsymbol{\wp}}_{m,\rm eff}|^2 \Gamma_0.
\end{align}
From this expression, one can then easily see that for transverse polarization only the mode with $m=0$ has a non-vanishing value of the effective dipole moment $\hat{\boldsymbol{\wp}}_{m,\rm eff}=\sqrt{N}\hat{e}_z$, and thus it is bright and decaying at rate $\Gamma_{m=0} \sim N\Gamma_0$. Instead for tangential or radial polarization there are two bright modes $m=\pm 1$ with $\hat{\boldsymbol{\wp}}_{m,\rm eff}=\sqrt{N/2} \left( \hat{e}_x \pm i \hat{e}_y \right)$ and $\Gamma_{m=\pm 1} \sim N\Gamma_0/2$. The remaining modes are dark with vanishing effective dipole moment and $\Gamma_m\rightarrow 0$. Figure \ref{Fig2}(c) shows the decay rates for a ring in this regime ($d/\lambda = 0.05$, $N=20$) with different polarization orientations $\dbu_i = \left\{\hat{e}_z, \hat{e}_{r,i}, \hat{e}_{\phi,i} \right\}$.  Moreover, note that in general, a ring with polarization  $\dbu_i = \cos\theta\cos\phi \, \hat{e}_{\phi,i} + \cos\theta\sin\phi \, \hat{e}_{r,i}+\sin\theta \,\hat{e}_z$ ($i=1,\dots, N$) will have three different bright modes $m = 0,\pm 1$ with decay rates $\Gamma_{m=0}=N\Gamma_0 \sin^2\theta$ and $\Gamma_{\pm 1} = (N\Gamma_0/2) \cos^2\theta$.

In this limit, the collective frequency shifts also acquire a particularly simple co-sinusoidal form. Indeed, in this regime the interactions $\Omega_{ij}$ between first neighbouring sites dominate, and one can approximate 
\[\Omega_m \approx N^{-1} \sum_{\ell} \left( \Omega_{\ell,\ell+1} e^{i 2\pi m /N} +\Omega_{\ell,\ell-1} e^{-i 2\pi m /N} \right) = 2\Omega_{d}\cos\left(2\pi m / N \right),\] where we use again the discrete rotational symmetry of the ring. Here the sign and strength of first-neighbour coupling $\Omega_d$ strongly depends on the polarization direction. For the same general polarization as before, $\Omega_d = - (3\Gamma_0/4k_0^3 d^3) \left[\cos^2\theta (3\cos^2\phi -\sin^2(\pi/N) )-1\right]$ \cite{cremer2020polarization}. Therefore, the bright modes will be energetically high (low) for transverse / radial (tangential) polarization, as it is shown in \fref{Fig2}(d) for the same parameters as before. Moreover, for polarization angles $\cos\theta \approx 1/\sqrt{3} \cos\phi$ and large number of emitters, a nearly degenerated flat band emerges, with frequency shifts that basically vanish \cite{cremer2020polarization}.

Finally, it is also possible to evaluate the electromagnetic field generated by one of these eigenmodes, by using Eq.\eqref{Eq:Fields}. The result will strongly depend on the angular momentum $m$, polarization orientation, and size of the ring. For the ring geometry, we find that strongly subradiant modes radiate with very low intensity basically along the ring plane, while the field is evanescent in the transverse direction, as shown in the top row of \fref{Fig9} and \fref{Fig10} for a ring of $N=9$ tangentially polarized emitters and $m=4$. Instead, the brightest modes (which in this case correspond to $m=\pm 1$) exhibit a strong field at the center of the ring and propagates also transversally to the ring plane, as shown in the same figures.

\subsection{Two Coupled Nanorings}

\begin{figure*}[t!]
\centering
\includegraphics[width=1\textwidth]{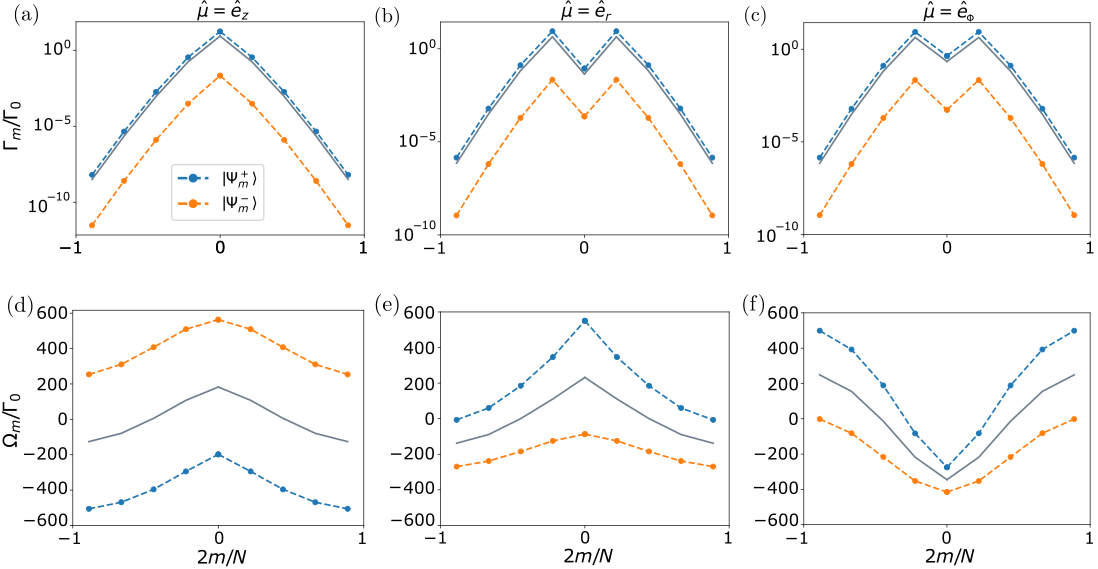}
\caption{{\bf Two coupled identical non-rotated nanorings ($\delta=0$)}. {\bf(a)}-{\bf(c)} Collective decay rates $\Gamma_m$ and {\bf(d)}-{\bf(f)} frequency shifts $\Omega_m$ versus angular momentum $m$, for two coupled rings of $N=9$ emitters each and $R/\lambda = 0.05$. The blue (orange) dashed lines denote the symmetric (anti-symmetric) eigenmodes. For comparison, the single ring solution for the same parameters is shown (grey solid line). The two rings are separated by the vertical distance $Z=0.5R$, and the emitters have transverse, radial or tangential polarization (left, middle or right panels, respectively). For transverse (radial and tangential) polarization the symmetric band is lower (higher) in energy.}
\label{Fig3}
\end{figure*}

We now analyze the case of two rings of radius $R_1$ and $R_2$ that are arranged concentrically and separated by a vertical distance $Z$. In general, we will also allow in the model a general rotation of angle $\delta \in [0,2\pi/N)$ of one of the rings around the $\hat{z}$-axis (see \fref{fig1}). In this case the unit cell consists of only two dipoles ($d=2$).\\

{\bf Coupled identical non-rotated rings} ($\delta = 0)$.-- We first focus on the case of two identical rings ($R_1 =  R_2$) concentrically stacked on top of each other and with no rotation angle $\delta$. Because the two rings are identical, and due to $\delta = 0$, the matrix $\tilde{G}_m^{\alpha \beta}$ is complex symmetric, and the eigenmodes of Eq.\eqref{HeffCoupled} can be chosen as the symmetric and the anti-symmetric superposition of Bloch states corresponding to each ring with well defined angular momentum $m$, which will be denoted as $\ket{\Psi^{\pm}_m} = (\ket{m,1} \pm \ket{m,2})/\sqrt{2}$ (with $\ket{m\alpha} \equiv \heg_{m\alpha} \ket{g}$). The corresponding collective frequency shifts and decay rates are then simply given by $\Omega_m^{\pm} = \Omega_m \mp \Omega^{\rm inter}_m$ and $\Gamma_m^\pm = \Gamma_m \pm \Gamma_m^{\rm inter}$, where $\Omega_m$ and $\Gamma_m$ are the frequency shift and decay rate corresponding to a single ring, whereas $\Omega_m^{\rm inter} = {\rm Re} [\tilde{G}_m^{12}]$ and
$\Gamma_m^{\rm inter} = -2{\rm Im} [\tilde{G}_m^{12}]$ are the dispersive and dissipative inter-ring couplings, respectively. 

In \fref{Fig3} we plot for two rings in the Dicke regime ($R / \lambda = 0.05$) and separated by vertical distance $Z=0.5 R$ the decay rates and frequency shifts of the two emerging bands: symmetric $\ket{\Psi^{+}_m}$ (orange line) and anti-symmetric $\ket{\Psi^{-}_m}$ (blue line). For comparison, we overlay the result for two independent rings (grey line). We find that, regardless of the emitters polarization, the anti-symmetric solution is always more subradiant than the symmetric one. Moreover the darkest state is $\Psi^-_{\max[m]}$, i.e., the anti-symmetric superposition of the darkest state of a single ring. Looking at the frequency shifts, we find that the behavior with angular momentum $m$ is similar to that for the single ring case, but shifted in energy. In particular, the symmetric band is shifted to lower energies (higher energies) for transverse (tangential and radial) polarization of the emitters. This fundamental difference in the energy shift sign can be intuitively understood in analogy to the energy of two interacting static dipoles. For the case with transverse polarization, two closer emitters from the two different rings are in a tail-to-head configuration, thus decreasing its total energy if they are in phase. Instead, for the case of tangential and radial polarization the emitters polarization is parallel, increasing its energy when they have the same phase. In conclusion, these results show that the polarization of the emitters can fundamentally modify the optical properties of the emerging bands and determine the ordering of states in energy, something which is relevant in the excitation transfer between the different energy bands. In particular the energy transfer in photosynthetic processes involving dipole interacting chromophores is understood via H- and J-aggregation. In J-aggregates, neighboring chromophores are oriented in a head-to-tail arrangement, resulting in a negative coherent nearest-neighbor coupling $\Omega_d$ and the positioning of the optically allowed ($m=0$) Bloch mode at the bottom of the energy band, whereas for H-aggregates the orientation is parallel and the symmetric ($m=0$) mode is positioned at the top of the energy band.
\begin{figure*}[t!]
\centering
\includegraphics[width=1\textwidth]{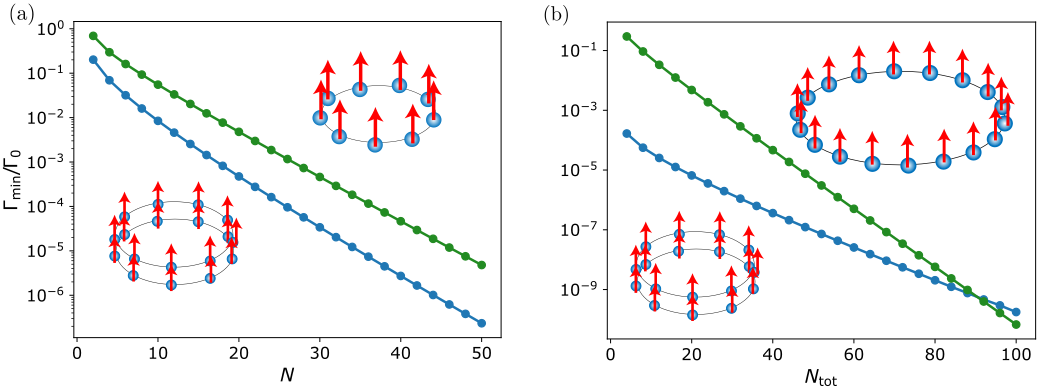}
\caption{{\bf Two coupled identical non-rotated nanorings ($\delta=0$)}. Scaling of the most subradiant eigenmode decay rate for two coupled rings (blue) with $Z/\lambda = 0.009$ versus the atom number $N$ of each of the rings. For comparison, we overlay the most subradiant decay rate for a single ring of $N$ atoms ({\bf a}) and a single ring of  $N_{\rm tot} = 2N$ atoms ({\bf b}) (green) with fixed inter-particle distance $d/\lambda = 1/3$ and transverse polarization. Similar results are found in case of tangential polarization.}
\label{Fig4}
\end{figure*}
Another interesting property of this system is the scaling of the most subradiant state decay rate with the atom number $N$. For a fixed inter-particle distance $d/\lambda$, we show in \fref{Fig4}(a) the decay rate of the most subradiant state of two coupled rings of $N$ emitters each, compared to that of a single ring of $N$ emitters (left panel). We observe that, in addition to a lower decay rate, the double ring structure shows always a stronger exponential suppression with the atom number compared to a single ring of the same size and interparticle distance $d$. In \fref{Fig4}(b) we also compare the double-ring result but with a single ring of $2N$ emitters and the same density. We find that in this case, for small inter-ring distances $z$ and ring atom number $N$, the coupling between the two rings is still strong enough to lead to more subradiance compared to the single ring case with the same total number of atoms. However, if $N$ is too large, then the single ring will always support the most subradiant state, as the curvature and therefore losses will experience a strong suppression as the system approaches an infinite linear chain, for which it is known that the decay rates are exactly zero. For this threshold  the exponential suppression with $N$ overcomes the coupling effect between the two rings. 
\begin{figure*}[t!]
\centering
\includegraphics[width=0.9\textwidth]{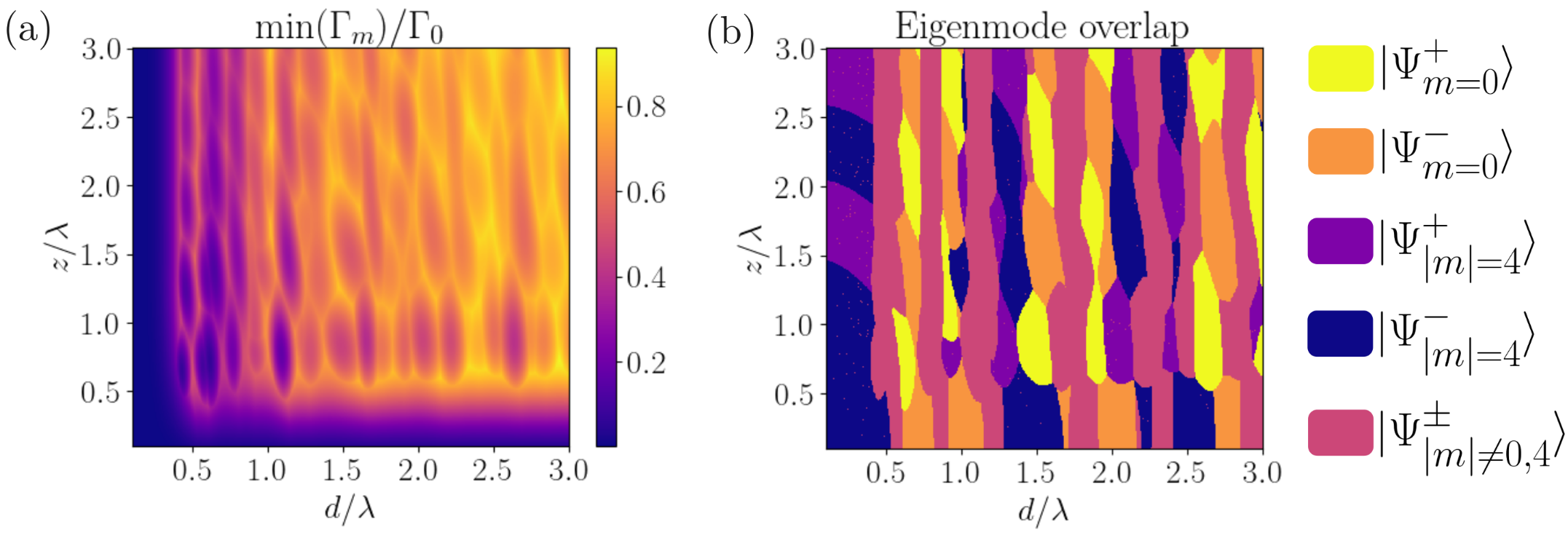}
\caption{{\bf Two coupled identical non-rotated nanorings ($\delta=0$)}. {\bf (a)} Most subradiant decay rate of two coupled rings with $N=9$ emitters and transverse polarization, as a function of ring constant $d/\lambda$ and inter-ring distance $z/\lambda$. Subradiant states can exist even beyond the threshold $d/\lambda < 1/2$ and $z/\lambda <1/2$ due to destructive wave interference. {\bf (b)} Overlap of the most subradiant eigenmode with the Bloch waves corresponding to angular momentum $|m|$. The Bloch waves of each ring can form symmetric and antisymmetric superpositions and it can be seen, that at various distances the symmetric superposition of $m=0$ Bloch waves can be subradiant.
The parameters are identical to (a) and the overlap oscillates when varying the ring parameters, as soon as $d,z \gtrsim \lambda/2$.}
\label{Fig5}
\end{figure*}
Interestingly, the most subradiant decay rate does not show a monotonic behavior with the lattice constant $d/\lambda$ or the inter-ring distance $z/\lambda$. In \ref{Fig5}(a) we plot the most subradiant decay rate versus these two ratios. We observe that the decay rate oscillates due to wave interference and that there can still exist subradiance beyond the values $d/\lambda=1/2$ and $z/\lambda = 1/2$. As previously discussed, such subradiant state is always the anti-symmetric superposition of two Bloch waves of well defined angular momentum $m$. For small rings such that $d/\lambda < 1/2$, the most subradiant state always corresponds to the superposition of the two most subradiant states, i.e., $|m| = \lceil (N-1)/2 \rceil$. However, for $d/\lambda > 1/2$ the value of $m$ that produces the most subradiant state periodically varies. This behavior is shown in \ref{Fig5}(b), where we have plotted the overlap of the Bloch waves of particular absolute value of the angular momentum. Additionally the Bloch waves can be in a symmetric or antisymmetric superposition and even the symmetric superposition of the symmetric $m=0$ modes can lead to subradiance at various distances.

\begin{figure*}[ht!]
\centering
\includegraphics[width=\textwidth]{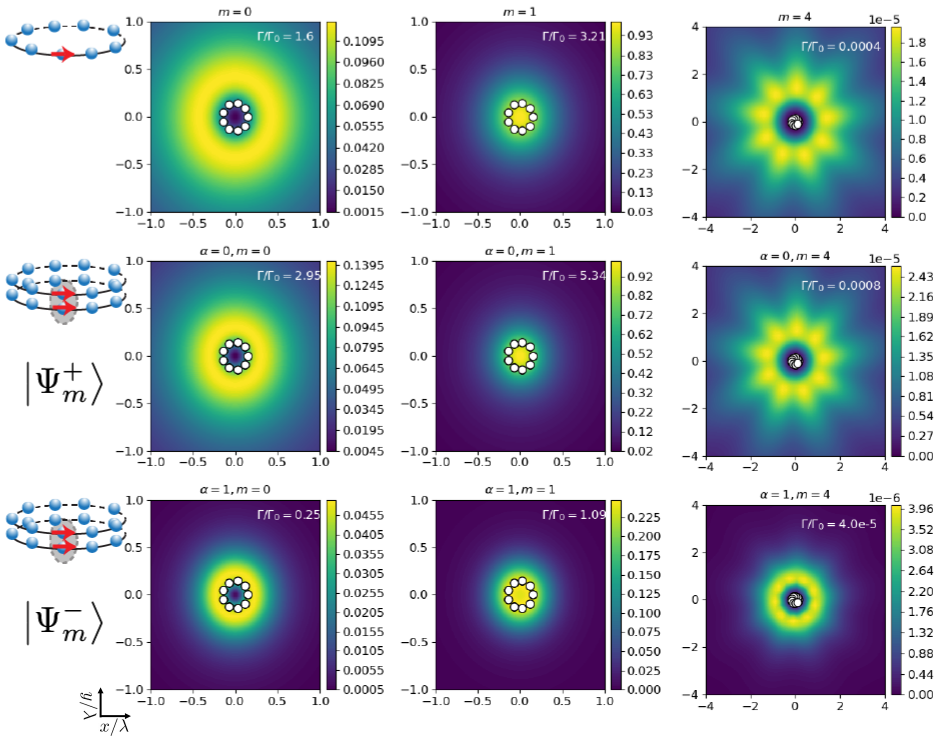}
\caption{{\bf Two coupled identical non-rotated nanorings ($\delta=0$).} Field intensity pattern versus real space coordinates in units of the transition wavelength $\lambda_0$ (cut at $z=6R$) generated by the eigenmodes with $m=0,1,4$ as indicated in the panels. Middle and bottom rows correspond to the symmetric and anti-symmetric eigenmodes, respectively. Top panels are for the single ring, for comparison. ($N=9$, $d/\lambda = 0.1$, $Z/\lambda = 0.2$, tangential polarization.)}
\label{Fig9}
\end{figure*}

\begin{figure*}[ht!]
\centering
\includegraphics[width=\textwidth]{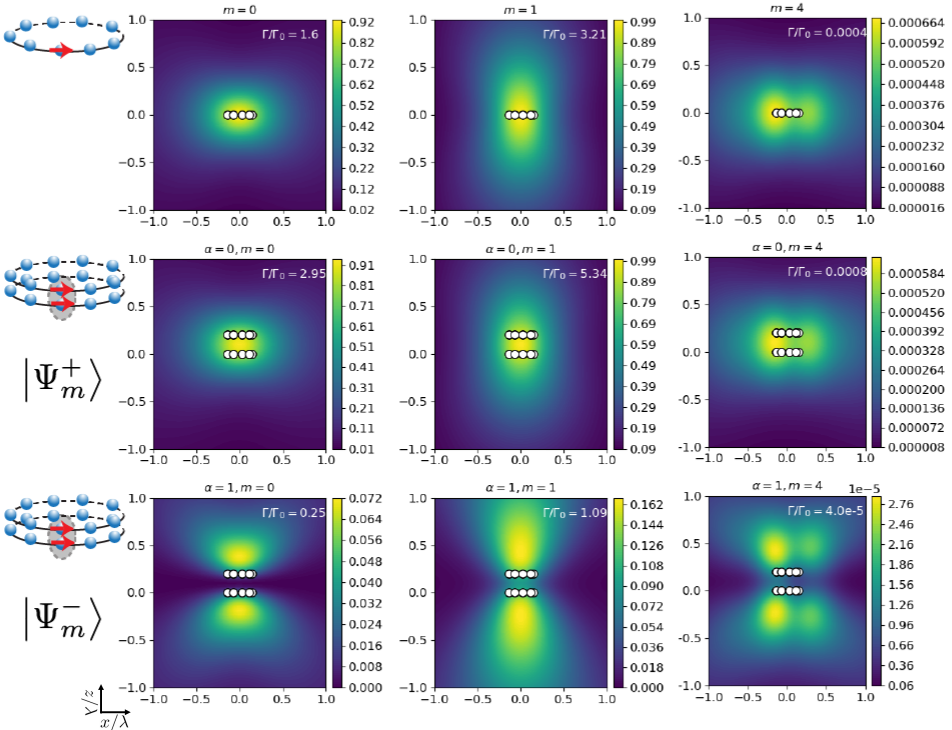}
\caption{{\bf Two coupled identical non-rotated nanorings ($\delta=0$).} Field intensity pattern versus real space coordinates (cut at $y=6R$) generated by the eigenmodes with $m=0,1,4$ as indicated in the panels. Middle and bottom rows correspond to the symmetric and anti-symmetric eigenmodes, respectively. Top panels are for the single ring, for comparison. ($N=9$, $d/\lambda = 0.1$, $Z/\lambda = 0.2$, tangential polarization.) }
\label{Fig10}
\end{figure*}

We finally discuss the striking differences in the field patterns generated by the eigenmodes $\ket{\Psi_m^\pm}$, with $m=0,1,4$. In \fref{Fig9} and \fref{Fig10} we plot (middle and bottom rows) the field intensity as a function of real space position, for two identical coupled concentric rings of $N=9$ emitters with tangential polarization, lattice constant $d/\lambda = 0.1$ and separated by a vertical distance $Z / \lambda =0.2$. For comparison, we have added in the top row the result for a single ring with the same parameters. We find that the symmetric superposition shows a pattern which is very similar to the single ring case. The brightest mode ($m=1$ in this case) shows an enhanced field intensity along the central axis of the rings. In the symmetric mode, the field is enhanced in the region between the two rings, whereas in the anti-symmetric superposition, it shows a striking different pattern with suppressed field in the region between the two rings.\\

{\bf Coupled unequal rings with rotation} ($\delta \neq 0$).--  We now consider the more general case where the two rings can have different radius and are rotated by an angle $\delta$. Note that in this case the matrix describing the single excitation manifold $\tilde{G}_m^{\alpha \beta}$ is in general not complex symmetric. However, for the equal radius case ($R_1 = R_2$) in the Dicke regime, the off-diagonal elements satisfy $\tilde{G}_m^{\alpha \beta} = (\tilde{G}_m^{\beta\alpha})^*$ ($\alpha \neq \beta$). This leads to eigenmodes of the form $\ket{\Psi^{\pm}_m} = (\ket{m,1} \pm e^{i \eta} \ket{m,2})/\sqrt{2}$ with $\eta = \textrm{atan} \left[\textrm{Im} \tilde{G}_m^{12}/  \textrm{Re} \tilde{G}_m^{12} \right]$. 

The behavior of the eigenmodes and eigenvalues with the rotation angle $\delta$ is not trivial and strongly depends on the polarization orientation and inter-particle distances. For transverse polarization and a small vertical separation between the rings ($Z = 0.1R$, $R/\lambda = 0.05$ we find a value of $\delta_c \sim 0.15$ for which the frequencies of the two eigenmodes with $m = \lceil (N-1)/2 \rceil$ feature an avoided level crossing. Interestingly, at this point the nature of the state changes. While for $\delta < \delta_c$ the highest energy state is radiant with $\eta \sim 0$, for $\delta > \delta_c$ the highest energy state becomes subradiant with $\eta \sim \pi$. These features are shown in \fref{Fig8} (top panels) and disappear for too small value of $Z$. Similar results can be found for other values of $m$. Moreover, the decay rate of the most subradiant state presents a broad minimum around $\pi/N$ and with $\eta \sim \pi/2$, i.e., when the sites of the second ring lie exactly in between those of the first ring. At this point and because the interparticle distances are larger, the frequency shifts are also smaller. 

Similar results can be found for other polarization orientations and also when varying the relative radius between the two rings. As an example, we show in \fref{Fig8} (bottom panels) the same analysis for two co-planar rings ($Z=0$) with tangential polarization and $R_1 = 0.9 R_2$. As it can be seen in the figure, in this case there is also an avoided level crossing (inset) at value $\delta_c \sim 0.07$, where the state of the highest energy state changes to be subradiant. As in the previous case, we find also the broad minimum around $\delta \sim \pi/N$, where the frequency shifts almost vanish. It is worth noting that in the natural light harvesting complex LH2 (see next section) the dipoles of the B850 band are arranged in a similar configuration with rotation angle $\delta \sim \pi/N$. An intriguing question is whether this is an accidental coincidence or whether the broad minimum emerging in the decay rate, which is thus robust against small fluctuations in the emitters position, can play a relevant role in the energy transfer and the light harvesting processes. 

\begin{figure*}[t!]
\centering
\includegraphics[width=1\textwidth]{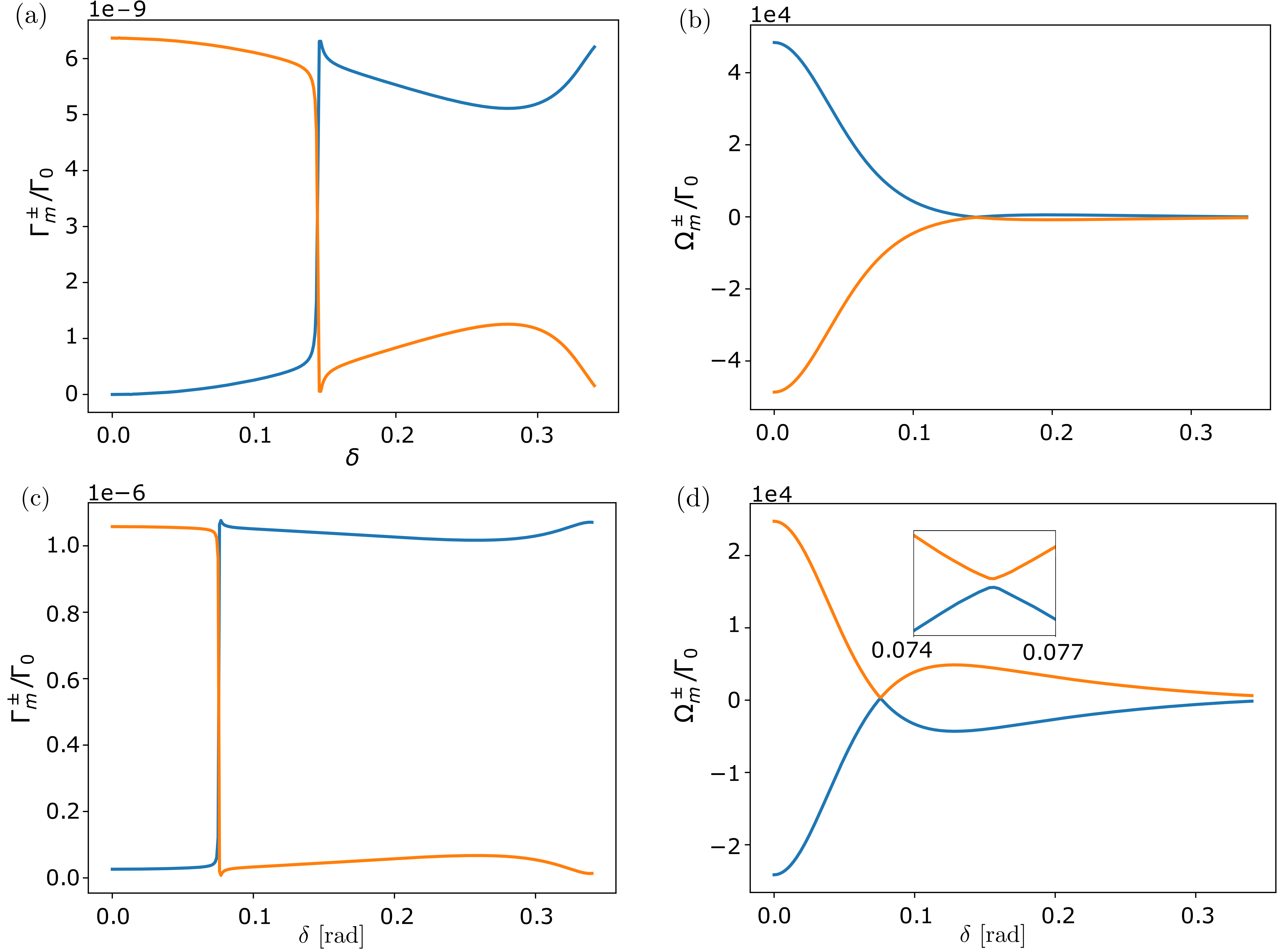}
\caption{{\bf Two coupled rotated nanorings} ($\delta \neq 0$). {(\bf Top panels)}  Two identical nanorings ($R = 0.05 \lambda$) with transverse dipole orientation separated by a vertical distance $Z = 0.1R$, depending on the rotation angle $\delta \in [0,2\pi/18]$: {\bf (a)} decay rate and {\bf (b)} frequency shift of the two eigenmodes with $m = \lceil (N-1)/2 \rceil$. %{\bf (c)} relative phase $\eta$ , 
An avoided level crossing emerges at $\delta \sim 0.15$, where the highest energy level changes from being subradiant to radiant, and from being  antisymmetric to symmetric. {(\bf Bottom panels)} Two coplanar unequal nanorings ($Z=0$) with radius $R_1 = 0.05 \lambda$ and $R_2 = 0.9 R_1$ and tangential dipole orientation, depending on the rotation angle $\delta$: {\bf (c)} decay rate and {\bf (d)} frequency shift of the two eigenmodes with $m = \lceil (N-1)/2 \rceil$. Similarly as before, an avoided level crossing (shown amplified in the inset) emerges at $\delta \sim 0.07$, where the highest energy level changes from being radiant to subradiant, and from being symmetric to anti-symmetric.}
\label{Fig8}
\end{figure*}

\clearpage
\subsection{B850 and B800 Bands in LH2} As already anticipated, the study of the optical properties of two (or more) coupled nanorings is motivated by the existence of similar structures in nature that enable efficient light harvesting and energy transfer \cite{bourne2019structure,cogdell2006architecture,mirkovic2017light,montemayor2018computational,caycedo2017quantum,pruchyathamkorn2020complex,jang2018delocalized,Fleming2000,Silbey2004,Cheng2006,Silbey2007,Olaya2008}. Indeed, while most biological systems are soft and disordered, photosynthetic complexes in certain purple bacteria exhibit crystalline order. The complexes are composed by antenna units that show a n-fold symmetry \cite{montemayor2018computational} which in turn, are arranged forming a maximally packed hexagonal pattern \cite{cleary2013optimal}. Purple bacteria are among the oldest living organisms and most efficient in turning sunlight into chemical usable energy. One of the most common species (Rhodopseudomonas Acidophila) contains two well differentiated types of complexes: a larger one containing the reaction center where the energy conversion takes place (LH1), and a second one (LH2) which is more abundant and whose main role is the absorption of photons and efficient subsequent energy transfer towards the LH1 units. The two complexes are formed by the same light-absorbing pigments: carotenoids (absorbing wavelengths ranging from $400$ to $550$ nanometers) and bacteriochlorophyl-a (BChla, absorbing in the red and infrared).  The BChla features a two-level optical dipole transition around $800-875$ nanometers (depending on the complex). These pigments are sustained by a hollow cylinder of apoproteins whose diameter is few tens of $\AA$.

\begin{figure*}
\centering
\includegraphics[width=1\textwidth]{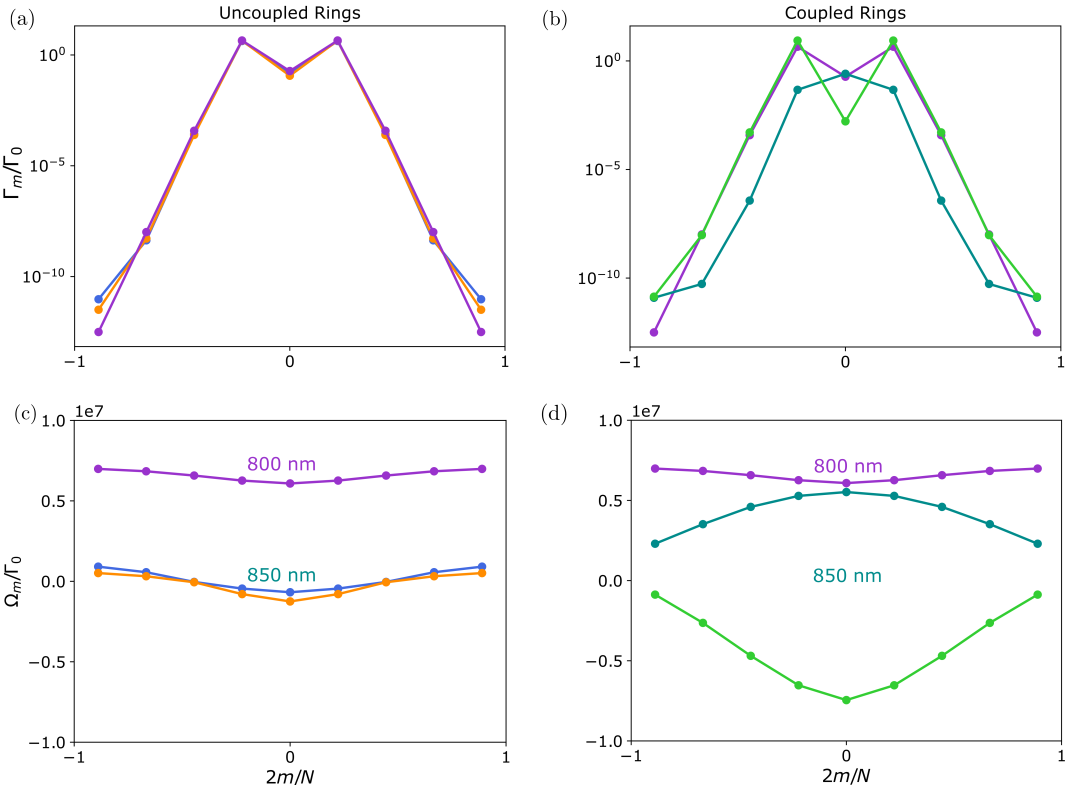}
\caption{{\bf LH2 dipole configuration}. {\bf (a)}-{\bf (b)} Collective decay rates and {\bf (c)}-{\bf (d)} frequency shifts as a function of angular momentum index $m$ for the LH2 structure (B800 and B850 bands) parameterized according to \cite{montemayor2018computational}. Left and Right panels correspond to uncoupled and coupled rings, respectively. The B850 band consists of a two-component unit cell ring with 9-fold symmetry (denoted by blue and orange), whereas the B800 band is a single component ring with 9-fold symmetry (denoted by violet). The B800 ring is far in energy and thus only couples very weakly to the B850 rings. However, the two components of the B850 band are strongly coupled, due to the reduced inter-particle distance, what leads to a broad dispersion in the frequency shifts. Two bands emerge: a darker band which is higher in energy and close to the B800 band, and a brighter band (with two bright modes corresponding to $m=\pm 1$) which is lower in energy. This band structure is relevant for the excitation energy transfer occurring between the B800 and B850 bands.}
\label{Fig6}
\end{figure*}

Here we will focus on the LH2 complex and the optical properties displayed by the BChla. Early x-ray crystallography data \cite{McDermott1995Crystal} together with subsequent molecular dynamics simulations \cite{Scholes1999} suggests a ring structure with 9-fold symmetry. This structure consists of a ring of $9$ emitters maximally absorbing at $800 \textrm{ nm}$ (the so-called $B800$ band) concentrically arranged and coupled to another two-component ring with 9-fold symmetry (with a total of $18$ emitters) maximally absorbing at $850 \textrm{ nm}$ (the so-called $B850$ band). The dipoles orientation also preserves the 9-fold rotational symmetry and are mostly contained in the plane of the ring, except for a small vertical component (see inset in \fref{Fig6}). Therefore, the whole structure can be regarded as a ring of $9-$unit cells of $3$ components (denoted by purple, blue and yellow in the figure). 

\begin{figure*}
\centering
\includegraphics[width=1\textwidth]{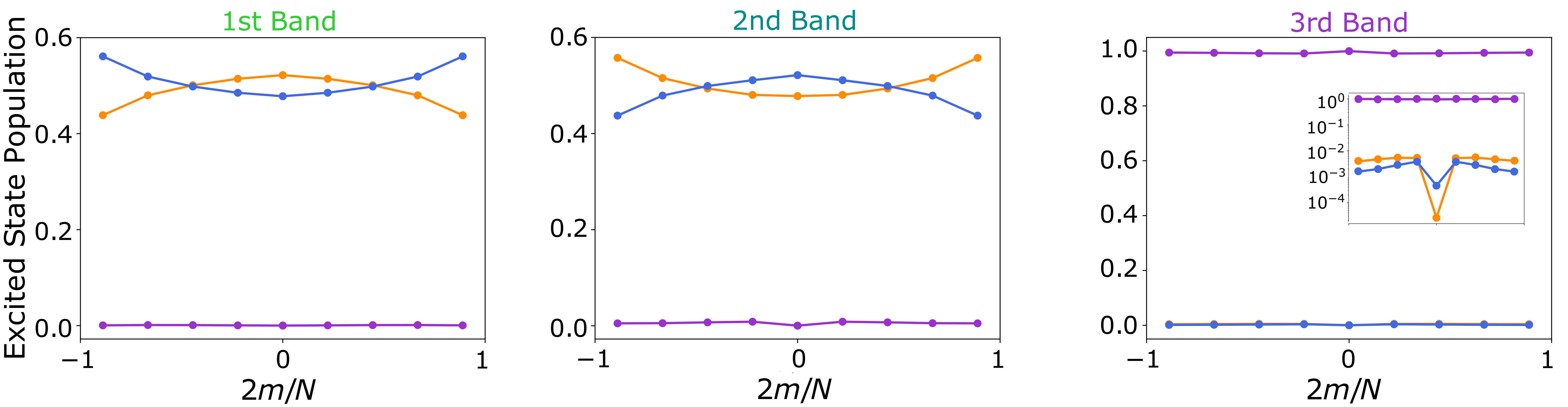}
\caption{{\bf LH2 dipole configuration}. Individual ring occupation propabilities for each of the three eigenmodes as a function of angular momentum index $m$. Blue and yellow correspond to the two B850 rings (as indicated in \fref{Fig6}), whereas violet is the occupation of the B800 ring. Each panel is a different eigenmode, indicated with the same code color as in \fref{Fig6}.}
\label{Fig7}
\end{figure*}

In the following, we analyse the eigenmodes and collective optical properties of the two bands ($B800$ and $B850$) using the parameters extracted from \cite{montemayor2018computational}. This analysis can be relevant for the understanding of the efficient energy transfer between the $B800$ and the $B850$ bands, but also for energy transfer between the LH2 and the LH1 units. Taking into account that the lifetime of the excited state in the BChla is of the order of nanoseconds, the energy transfer process is expected to occur at a much faster time scale. Figure \ref{Fig6} shows and compares the decay rates and frequency shifts of the collective eigenmodes as a function of the angular momentum quantum number $m$, considering the rings are uncoupled (left) or coupled (right). The dispersive couplings between the two components of the $B850$ band (denoted by yellow and blue in the figure) are very large due to the small inter-particle distances, and of the order of $10^6 \Gamma$ (being $\Gamma \sim 25 \textrm{ MHz}$ the estimated decay rate of the excited state in the dipole transition). This leads to the emergence of a two-band structure with large  frequency splitting where the two components of the $B850$ ring strongly hybridize: a higher energy band which is mostly subradiant, and a lower energy band containing only two bright modes at $m=\pm 1$. For completeness, we show in \fref{Fig7} the excited state population of each of the components for the coupled system eigenmodes. Clearly, the excitation is delocalized over the two components of the ring.

In the inset of \fref{Fig7} we show the small contributions of the lower double ring configuration to the excited state population of the third band. A similar behavior emerges in case of the first and second band, where the $B800$ ring gives a non-vanishing contribution to the population of the first and second band.

In contrast, the coupling between the $B850$ and $B800$ band (indicated by purple in the plot) is ten times smaller (of the order $10^5 \Gamma$), whereas the energy transition difference is of the order of $10^7 \Gamma$, and therefore, the $B800$ band remains mostly decoupled. However, it is worth noting that after the $B850$ bands are coupled, the higher energy band lies close to the $B800$ band.

Finally let us point out a very special property of the naturally occurring geometry. Indeed it can be seen that the actual geometry is very close to the critical transition point, where the up shifted eigenstate energies of the lower double ring just overlap with the upper ring energies. For this in Fig.~\ref{Fig8} we plot the corresponding exciton energies as function of the overall size of the molecule, where we only consider small size variations $R_{\alpha,i} = \alpha R_i$ around the actually measured size. We see that closely below the value of $\alpha = 1$ the energy bands cross and eigenstates appear which posses similar contributions of all three rings. Close to this resonance condition any excitation in one of the rings is thus coherently transported to the other rings in short time. Interestingly the crossing point depends on the angular index $m$ shifting further away from $\alpha = 1$ with growing $m$. From this sensity behaviour one could expect tune-ability of the ring properties via the local refractive index or small deformations of the complex.   
\begin{figure*}
\centering
\includegraphics[width=1\textwidth]{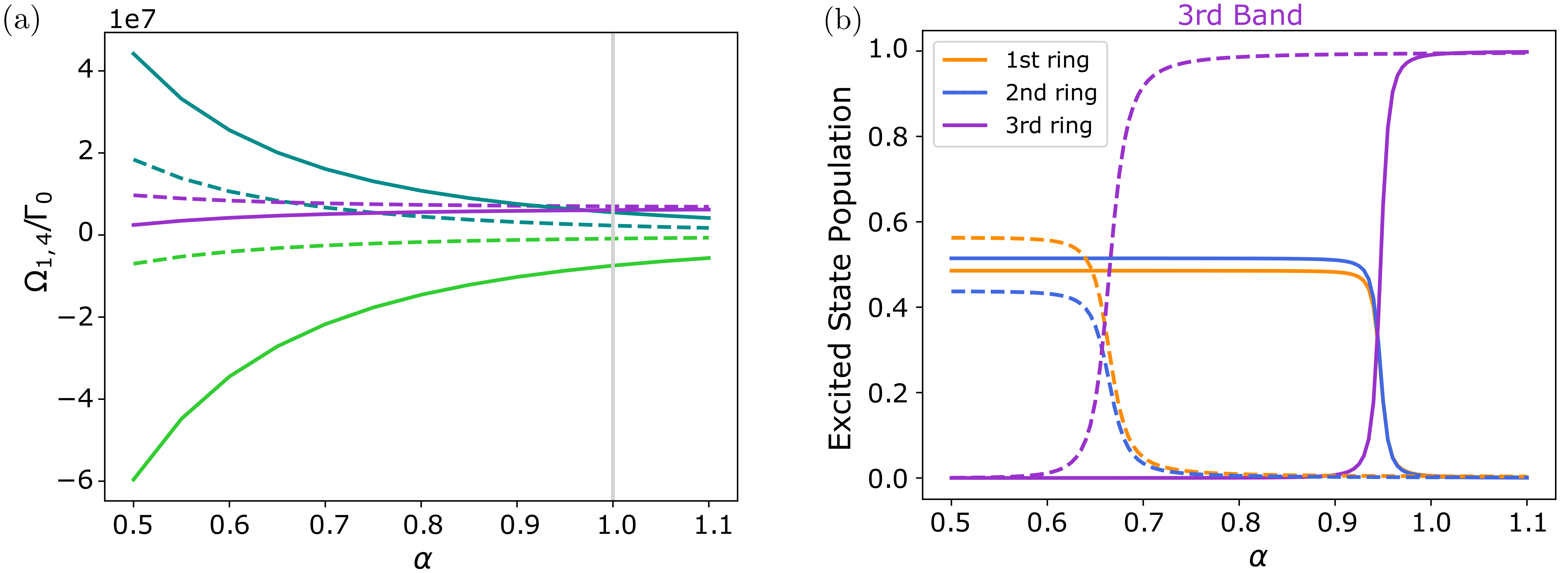}%
\caption{{\bf LH2 dipole configuration}. {\bf (a)} Frequency shift and {\bf (b)} ring occupation probabilities for the third band as function of the overall size of the molecule at the $m=1,4$ mode. The size of ring $i$ is varied via $R_{\alpha,i} = \alpha R_i$ around the actual size ($\alpha=1$). Solid (dashed) lines correspond to the $m=1$ ($m=4$) mode. The color code is equivalent to the one in Fig.~\ref{Fig6}. Dependent on the mode $m$ the second and third band as well as the excited state populations cross at $\alpha_\mathrm{c} < 1$. For systems with $\alpha < \alpha_c$, the third band is occupied by the B850 ring whereas for $\alpha > \alpha_c$ it is occupied by the B800 ring.}
\label{Fig8}
\end{figure*}

\section{Conclusions and Outlook}
Our calculations show that structures involving multiple concentric rings exhibit strongly modified exciton properties and in particular, feature extremely subradiant states with sub-wavelength confined fields. For two identical rings at close enough distances, we find that the anti-symmetric superposition of the individual ring radiative modes, which inherits the angular symmetry of the setup, is always more subradiant than the corresponding symmetric combination. In particular, the most subradiant states are obtained by choosing the individual ring darkest eigenmodes. We have shown that the spontaneous emission of such states decays faster with the emitter number compared to the single ring case. 
 
Moreover, important radiative properties, such as the ordering in frequency of the optical modes,  can be controlled via a relative rotation or size differences of two otherwise identical rings. For instance, we find that by modifying these parameters, the highest energy level changes from being subradiant to superradiant.

When we apply our model to the specific geometry of the triple ring LH2 structure including the natural distances, energy shifts and dipole polarization, we find most of the collective modes are extremely dark. Most interestingly the collective energy shifts from the lower double B850 ring structure, for which the interparticle distances are very small, is of the order of the $50$ nm energy shift of the upper ring, so that the energy spectrum spans almost the full gap between the rings. %gets close for the grey $m=0$ symmetric states. 
More specifically, two bands emerge due to the strong coupling between the two B850 components. A subradiant band which is higher in energy and close to the B800 band, and a brighter band which is much lower in energy. The realistic dipole orientations and distances lead to only two bright modes corresponding to a quasi-symmetric superposition of the angular momentum $m=1$ and $m=-1$ modes. This emerging band structure could be helpful for any phonon induced collective energy transfer processes, which are, of course, beyond our model here, but which we plan to explore in future work.

\clearpage

\end{document}